# Learning Diffractive Optical Communication Around Arbitrary Opaque Occlusions

Md Sadman Sakib Rahman[1,2,3], Tianyi Gan[1,3], Emir Arda Deger[1], Çağatay Işıl[1,2,3], Mona Jarrahi[1,3], and Aydogan Ozcan[1,2,3]

[1]Electrical and Computer Engineering Department, University of California, Los Angeles, CA, 90095, USA
[2]Bioengineering Department, University of California, Los Angeles, CA, 90095, USA
[3]California NanoSystems Institute (CNSI), University of California, Los Angeles, CA, 90095, USA
[*]Corresponding author: ozcan@ucla.edu

## Abstract

Free-space optical systems are emerging for high data rate communication and transfer of information in indoor and outdoor settings. However, free-space optical communication becomes challenging when an occlusion blocks the light path. Here, we demonstrate, for the first time, a direct communication scheme, passing optical information around a fully opaque, arbitrarily shaped obstacle that partially or entirely occludes the transmitter's field-of-view. In this scheme, an electronic neural network encoder and a diffractive optical network decoder are jointly trained using deep learning to transfer the optical information or message of interest around the opaque occlusion of an arbitrary shape. The diffractive decoder comprises successive spatially-engineered passive surfaces that process optical information through light-matter interactions. Following its training, the encoder-decoder pair can communicate any arbitrary optical information around opaque occlusions, where information decoding occurs at the speed of light propagation. For occlusions that change their size and/or shape as a function of time, the encoder neural network can be retrained to successfully communicate with the existing diffractive decoder, without changing the physical layer(s) already deployed. We also validate this framework experimentally in the terahertz spectrum using a 3D-printed diffractive decoder to communicate around a fully opaque occlusion. Scalable for operation in any wavelength regime, this scheme could be particularly useful in emerging high data-rate free-space communication systems.

## Introduction

Traditionally radio frequency (RF) and microwave have dominated the area of wireless communication. To meet the growing need for faster data transfer rates, RF systems employ increasingly complex coding, multiple antennas, and higher carrier frequencies (*1*). For example, by utilizing higher frequency bands, 6$^{th}$ generation (6G) technology is predicted to provide 100 to 1000 times faster speed than 5$^{th}$ generation (5G) systems deployed for wireless communication (*2*). With ever-increasing data rates, maintaining the performance of these schemes will become more challenging. One possible solution is to shift to shorter wavelengths, such as the ultraviolet (UV), visible or infrared (IR) regions of the electromagnetic spectrum, which provide much wider bandwidths compared to radio waves or microwaves (*1*, *3*–*5*). However, free-space optical communication becomes challenging when opaque occlusions block the light path. Non-line-of-sight (NLOS) communication, which exploits diffusely reflected waves from a nearby scattering medium, has been used as a way around the occlusion problem (*6*–*10*). However, the adaptability of these solutions to emerging optical communication techniques for channel capacity expansion faces challenges since even weak turbulence can cause a significant loss of information (*10*). Furthermore, the low power efficiency arising from the weak scattering or diffuse reflection is another limitation of NLOS communication. Other NLOS systems, e.g., for imaging around corners, also exist (*11*–*24*); these approaches, however, involve relatively slow and power-consuming digital methods for image reconstruction. Alternative methods have been developed for image transmission through thick (but transmitting) occlusions, including e.g., holography (*25*–*27*), adaptive wavefront control (*28*–*30*), and others (*31*, *32*). However, many of these techniques also involve digital reconstruction of the information, often requiring iterative algorithms. Moreover, these are applicable for multiple scattering media that are transmissive, and do not address situations, where the light path is either partially or entirely obstructed by opaque occlusions with zero light transmittance.

Here we demonstrate a novel scheme for directly communicating optical information of interest around zero-transmittance occlusions using electronic encoding at the transmitter and all-optical diffractive decoding at the receiver. In our scheme, an electronic neural network, trained in unison with an all-optical diffractive decoder, encodes the message of interest to effectively bypass the opaque occlusion and be decoded at the receiver by an all-optical decoder, using passive diffraction through thin structured layers. This all-optical decoding is performed on the encoded wavefront that carries the optical information or the message of interest, after its obstruction by an arbitrarily shaped opaque occlusion. The diffractive decoder processes the secondary waves scattered through the edges of the opaque occlusion using a passive, smart material comprised of successive spatially engineered surfaces,(*33*) and performs the reconstruction of the hidden information at the speed of light propagation through a thin diffractive volume that axially spans < $100\times\lambda$, where $\lambda$ is the wavelength of the illumination light.

We show that this combination of electronic encoding and all-optical decoding is capable of direct optical communication between the transmitter and the receiver even when the opaque occlusion body entirely blocks the transmitter's field-of-view (FOV). We also report an experimental demonstration of this scheme using a 3D-printed diffractive decoder that operates at the terahertz spectrum. Furthermore, we demonstrate that this scheme could be configured to be highly power efficient, reaching diffraction efficiencies of >50% at its output. In the case of opaque occlusions that change their size/shape over time, we also report that the encoder neural network could be retrained to successfully communicate with an existing diffractive decoder, without changing its physical structure that is already deployed. This makes the presented concept highly dynamic and easy to adapt to external and uncontrolled changes that might happen between the transmitter and receiver apertures. This framework can be extended for operation at different parts of the electromagnetic spectrum, and would find applications in emerging high-data-rate free-space communication technologies, under scenarios where different undesired structures occlude the direct channel of communication between the transmitter and the receiver.

## Results

A schematic depicting the optical communication scheme around an opaque occlusion with zero light transmittance is shown in Fig. 1a. The message to be transmitted, e.g., the image of an object, is fed to an electronic/digital neural network, which outputs a phase-encoded optical representation of the message. This code is imparted onto the phase of a plane-wave illumination, which is transmitted toward the decoder through an aperture that is partially or entirely blocked by an opaque occlusion. The scattered waves from the edges of the opaque occlusion travel toward the receiver aperture as secondary waves, where a diffractive decoder all-optically decodes the received light to directly reproduce the message/object at its output FOV. This decoding operation is completed as the light propagates through the thin decoder layers. For this collaborative encoding-decoding scheme, the electronic encoder neural network and the diffractive decoder are jointly trained in a data-driven manner for effective optical communication, bypassing the fully opaque occlusion positioned between the transmitter aperture and the receiver.

Figures 1b and 1c provide a deeper look into the encoder and the decoder architectures used in this work. As shown in Fig. 1b, the convolutional neural network (CNN) encoder is composed of several convolution layers, followed by a dense layer representing the encoded output. This dense layer output is rearranged into a 2D-array corresponding to the spatial grid that maps the phase-encoded transmitter aperture. We assumed that both the desired messages and the phase codes to be transmitted comprise $28 \times 28$ pixels unless otherwise stated. The architecture of the encoder remains the same across all the designs reported in this paper. The architecture of the diffractive decoder, which decodes the transmitted and obstructed phase-encoded waves, is shown in Fig. 1c. This figure shows a diffractive decoder comprising $L = 3$ spatially-engineered surfaces/layers (i.e., $S_1$, $S_2$ and $S_3$); however, in this work, we also report results for designs comprising diffractive decoders with $L = 1$ and $L = 5$ layers, used for comparison. Together with the encoder CNN parameters, the spatial features of the diffractive surfaces of the all-optical decoder are optimized to decode the encoded and blocked/obscured wavefront. In this work, we consider *phase-only* diffractive features, i.e., only the phase values of the features at each diffractive surface are trainable (see the 'Materials and Methods' section for details). Figure 1 also compares the performance of the presented electronic encoding and diffractive decoding scheme to that of a lens-based camera. As shown in Fig. 1d, the lens images reveal significant loss of information caused by the opaque occlusion in a standard camera system, showcasing the scale of the problem that is addressed through our proposed approach.

For all the models reported in this work, the data-driven joint training of the electronic encoder CNN and the diffractive decoder was accomplished by minimizing a structural loss function defined between the object (ground-truth message) and the diffractive decoder output, using 55,000 images of handwritten digits from the MNIST (*34*) training dataset, augmented by 55,000 additional custom-generated images (see the 'Materials and Methods' section as well as Supplementary Fig. S1 for details). All our results come from blind testing with objects/messages never used during training.

## Numerical analysis of diffractive optical communication around opaque occlusions

First, we compare, for various levels of opaque occlusions, the performance of trained encoder-decoder pairs with different diffractive decoder architectures in terms of the number of diffractive surfaces employed. Specifically, for each of the occlusion width values, i.e., $w_o = 32.0\lambda$, $w_o = 53.3\lambda$ and $w_o = 74.7\lambda$, we designed three encoder-decoder pairs, with $L = 1$, $L = 3$, and $L = 5$ diffractive layers within the decoders, and compared the performance of these designs for new handwritten digits in Fig. 2. This blind testing refers to 'internal generalization' because even though these particular test objects were never used in training, they are from the same dataset. As shown in Fig. 2, even $L = 1$ designs can faithfully decode the message for optical communication around these various levels of occlusions. Furthermore, as the number of layers in the decoder increases to $L = 3$ or $L = 5$, the quality of the output also gets better. While the performance of the $L = 1$ design deteriorates slightly as $w_o$ increases, the $L = 3$ and $L = 5$ designs do not show any appreciable degradation in qualitative performance for such bigger occlusions. Note that the width of the transmitter aperture is $w_t = 59.73\lambda$; therefore, for an occlusion size of for $w_o = 74.7\lambda$, none of the ballistic photons can reach the receiver aperture since the opaque occlusion completely blocks the aperture of the encoding transmitter aperture. Nonetheless, the scattering from the occlusion edges suffices for the encoder-decoder pair to communicate faithfully.

To supplement the qualitative results of Fig. 2, we also quantified the performance of different encoder-decoder pairs designed for increasing occlusion widths ($w_o$), in terms of peak signal-to-noise ratio (PSNR) and structural similarity index measure (SSIM) (*35*) averaged over 10,000 handwritten digits from the MNIST test set (never used before); see Figs. 3a and 3b, respectively. With increasing $w_o$, we see a larger decrease in the performance of $L = 1$ designs compared to $L = 3$ and $L = 5$ designs. Interestingly, there is a slight improvement in the performance of $L = 1$ and $L = 3$ decoders as $w_o$ surpasses $w_t = 59.73\lambda$ (the transmitter aperture width); this improved level of performance is retained for $w_o > w_t$, the cause of which will be discussed later in our Discussion section.

Next, for the same designs reported in Fig. 2, we explored the *external* generalization of these encoder-decoder pairs by testing their performance on types of objects that were not represented in the training set; see Fig. 4. For this analysis, we randomly chose two images of fashion products from the Fashion-MNIST (*36*) test set (top) and two additional images from the CIFAR-10 (*37*) test set (bottom). As shown in Fig. 4, our encoder-decoder designs show excellent generalization to these completely different object types. Although the decoder outputs of the $L = 1$ decoder designs for $w_o = 53.3\lambda$ and $w_o = 74.7\lambda$ are slightly degraded, the objects are still recognizable at the output plane even for the complete blockage of the transmitter aperture by the occlusion.

We also investigated the ability of these designs to resolve closely separated features in their outputs. For this purpose, we transmitted test patterns consisting of four closely spaced dots, and the corresponding diffractive decoder outputs are shown in Fig. 5. For the top (bottom) pattern, the vertical/horizontal separation between the inner edges of the dots is $2.12\lambda$ ($4.24\lambda$). None of the designs could resolve the dots separated by $2.12\lambda$; however, the dots separated by $4.24\lambda$ were resolved by all the encoder-decoder designs with good contrast, as can be seen from the cross-sections accompanying the output images in Fig. 5. It is to be noted that this resolution limit of $4.24\lambda$ is due to the output pixel size, which was set as $2.12\lambda$ in our simulations. The effective resolution of our encoder-decoder system can be further improved within the diffraction limit of light by using higher-resolution objects and a smaller pixel size during the training.

## Impact of phase bit depth on performance

Here, we study the effect of a finite bit-depth $b_q$ phase quantization of the encoder plane as well as the diffractive layers. For the results presented so far, we did not assume either to be quantized, i.e., an infinite bit-depth of phase quantization was assumed. For the $w_o = 32.0\lambda$, $L = 3$ design (trained assuming an infinite bit-depth $b_{q,tr} = \infty$), the first row of Fig. 6a shows the impact of quantizing the encoded phase patterns as well as the diffractive layer phase values with a finite bit-depth $b_{q,te}$. This represents an "attack" on the design since the encoder CNN and the diffractive decoder were trained without such a phase bit-depth restriction; stated differently, they were trained with $b_{q,tr} = \infty$ and are now tested with finite levels of $b_{q,te}$. For the $b_{q,tr} = \infty$ designs, the output quality remains unaffected for $b_{q,te} = 8$; however, there is considerable degradation under $b_{q,te} = 4$, and we face complete failure with $b_{q,te} = 3$ and $b_{q,te} = 2$. However, this sharp performance degradation with decreasing $b_{q,te}$ can be amended by considering the finite bit-depth during training. To showcase this, we trained two additional designs with $w_o = 32.0\lambda$ and $L = 3$ assuming finite bit-depths of $b_{q,tr} = 4$ and $b_{q,tr} = 3$; their blind testing performance with decreasing $b_{q,te}$ is reported in the second and third rows of Fig. 6a, respectively. Both of these designs show robustness against bit-depth reduction up to $b_{q,te} = 3$ (i.e., 8-level phase quantization at the encoder and decoder layers). However, even with $b_{q,te} = 2$ (only 4-level phase quantization), the outputs are still recognizable as shown in Fig. 6. We also quantified the performance (PSNR and SSIM) of these three designs ($b_{q,tr} = \infty$ $b_{q,tr} = 4$, $b_{q,tr} = 3$) for different $b_{q,te}$ levels; see Figs. 6b and 6c. These quantitative comparisons restate the same conclusion: training with a lower $b_{q,tr}$ results in robust encoder-decoder designs that preserve their optical communication quality despite a reduction in the bit-depth $b_{q,te}$, albeit with a relatively small sacrifice in the output performance.

## Output power efficiency

Next, we investigate the power efficiency of the optical communication scheme around opaque occlusions using jointly-trained electronic encoder-diffractive decoder pairs. For this analysis, we defined the diffraction efficiency (DE) as the ratio of the optical power at the output FOV to the optical power departing the transmitter aperture. In Fig. 7a, we plot the diffraction efficiency of the same designs shown in Fig. 3, as a function of the occlusion size. These values are calculated by averaging over 10,000 MNIST test images. These results reveal that the diffraction efficiency decreases monotonically with increasing occlusion width, as expected. Moreover, the diffraction efficiencies are relatively low, i.e., below or around 1%, even for small occlusions. However, this issue of low diffraction efficiency can be addressed in the design stage by adding to the training loss function an additional loss term that penalizes low diffraction efficiency (see the Supplementary Materials). Figure 7b depicts the improvement of diffraction efficiency resulting from increasing the weight ($\eta$) of this additive loss term during the training stage. For example, the $\eta = 0.02$ and $\eta = 0.1$ designs yield an average diffraction efficiency of 27.43% and 52.52%, respectively, while still being able to resolve various features of the target images as shown in Fig. 7c. This additive loss weight $\eta$ therefore provides a powerful mechanism for improving the output diffraction efficiency significantly with a relatively small sacrifice in the image quality as exemplified in Figs. 7b-c.

## Occlusion shape

So far, we have considered square-shaped opaque occlusions placed symmetrically around the optical axis. However, our proposed encoder-decoder approach is not limited to square-shaped occlusions and, in fact, can be used to communicate around any arbitrary occlusion shape. In Fig. 8, we show the performance comparison of four different trained encoder-decoder pairs for four different occlusion shapes, where the areas of the opaque occlusions were kept approximately the same. We can see that the shape of the occlusion does not have any perceptible effect on the output image quality. We also plot the average SSIM values calculated for these four models over 10,000 MNIST test images (internal generalization) as well as 10,000 Fashion-MNIST test images (external generalization) in Supplementary Fig. S2, which further confirm the success of our approach for different occlusion structures, including randomly shaped occlusions as shown in Fig. 8e.

## Experimental validation

We experimentally validated the electronic encoding-diffractive decoding scheme for communication around opaque occlusion in the terahertz (THz) part of the spectrum ($\lambda = 0.75\text{mm}$) using a 3D-printed single-layer ($L = 1$) diffractive decoder (see the 'Materials and Methods' section for details). We depict the setup used for this experimental validation in Fig. 9a. Figures 9b and 9c show the 3D printed components used to implement the encoded (phase) patterns, the opaque occlusion, and the diffractive decoder layer. Shown in Fig. 9c, the width of the transmitter aperture (dashed red square) housing the encoded phase patterns was selected as $w_t \approx 59.73\lambda$, whereas the width of the opaque occlusion (dashed green square) was $w_o \approx 32.0\lambda$ and the diffractive decoder layer (dashed blue square) width was selected as $w_l \approx 106.67\lambda$. The axial distances between the encoded object and the occlusion, between the occlusion and the diffractive layer, and the diffractive layer and the output FOV were $\sim 13.33\lambda$, $\sim 106.67\lambda$, and $\sim 40\lambda$, respectively. In Fig. 9d, we show the input objects/messages, the simulated lens images, and the simulated and experimental diffractive decoder output images for ten different handwritten digits randomly chosen from the test dataset. Our experimental results reveal that the CNN-based phase encoding followed by diffractive decoding resulted in successful communication of the intended objects/messages around the opaque occlusion (see the bottom row of Fig. 9d).

## Discussion

Our optical communication scheme using CNN-based encoding and diffractive all-optical decoding would be useful for the optical communication of information around opaque occlusions caused by existing or evolving structures. In case such occlusions change moderately over time (for example grow in size as a function of time), the same diffractive decoder that is deployed as part of our communication link can still be used with only an update of the digital encoder CNN. To showcase this, in Supplementary Fig. S3, we illustrate an encoder-decoder design with $L = 3$ that was originally trained with an occlusion size of $w_o = 32.0\lambda$ (blue boxes), successfully communicating the input messages between the CNN-based phase transmitter aperture and the output FOV of the diffractive decoder when the occlusion size remains the same, i.e., $w_o = 32.0\lambda$ (dashed blue box). The same figure also illustrates the failure of this encoder-decode pair once the size of the opaque occlusion grows to $w_o = 40.0\lambda$ (dotted blue box); this failure due to the (unexpectedly) increased occlusion size can be repaired without changing the deployed diffractive decoder layers by just retraining the CNN encoder part; see Supplementary Fig. S3, dashed green box.

The speed of optical communication through our encoder-decoder pair would be limited by the rate at which the encoded phase patterns (CNN outputs) can be refreshed or by the speed of the output detector-array, whichever is smaller. The transmission and the decoding processes of the desired optical information/message occur at the speed of light propagation through thin diffractive layers and do not consume any external power (except for the illumination light). Therefore, the main power consuming steps in our architecture are the CNN inference, the transmitter of the encoded phase patterns and the detector-array operation.

The communication around occlusions using our scheme works even when the occlusion width is larger than the width of the transmitter aperture since it utilizes CNN-based phase encoding of information to effectively exploit the scattering from the edges of the occlusions. Surprisingly, as the occlusion width surpasses the transmitter aperture width ($w_t$), the performance of $L = 1$ and $L = 3$ designs slightly improved, as was seen in Fig. 3. This relative improvement might be explained by a switch in the mode of operation of our encoder-decoder pair. When the opaque occlusions are smaller than the transmitter aperture, the pixels at the edges of the transmitter can communicate directly to the receiver aperture and therefore, they dominate the power balance. In this operation regime, as the occlusion size gets larger, the effective number of pixels at the transmitter aperture that directly communicates with the receiver/decoder gets smaller, causing a decline in the performance of the diffractive decoder. However, when the occlusion becomes larger than the transmitter aperture, none of the input pixels can dominate the power balance at the receiver end by communicating with it directly; instead, all the pixels of the encoder plane are forced to *indirectly* contribute to the receiver aperture through the edge scattering of the occlusion. This causes the performance to get better for occlusions larger than the transmitter aperture since effectively more pixels of the encoder plane can contribute to the receiver aperture without a major power imbalance among these secondary wave-based contributions (through edge scattering). This turnaround in performance (i.e., the switching behavior between these two modes of operation) is not observed when the diffractive decoder has a deeper architecture (e.g., $L = 5$) since deeper decoders can effectively balance the ballistic photons that are transmitted from the edge pixels; consequently, edge-pixels of the transmitter aperture do not dominate the output signals even when they can directly 'see' the receiver aperture since multiple layers of a deeper diffractive decoder act as a universal mode processor (*38–41*).

Finally, the success of the simpler decoder designs with $L = 1$ layer, as shown in Figs. 2-5 and 9, begs the question of whether such an optical communication around opaque occlusions is also feasible with electronic encoding only, i.e., without diffractive decoding. To address this question, we trained two encoder-only designs, for $w_o = 32.0\lambda$ and $w_o = 53.3\lambda$, and compared their performance against $L = 1$ designs in Supplementary Fig. S4. The encoder-only architecture barely succeeds for $w_o = 32.0\lambda$ and fails drastically for $w_o = 53.3\lambda$, whereas $L = 1$ designs provide significantly better performance. This demonstrates the importance of complementing electronic encoding with diffractive decoding for effective communication around opaque occlusions.

# Materials and Methods

## Model

In our model, the message/object $m$ that is to be transmitted is fed to a CNN, which yields a phase-encoded representation $\psi$ of the message. The message is assumed to be in the form of an $N_{in} \times N_{in} = 28 \times 28$ pixel image. The coded phase $\psi$ is assumed to have dimension $N_{out} \times N_{out} = 28 \times 28$. The $N_{out} \times N_{out}$ phase elements are distributed over the transmitter aperture of area $w_t \times w_t$, where $w_t \approx 59.73\lambda$ and $\lambda$ is the illumination wavelength. The lateral width of each phase element/pixel is therefore $w_t/N_{out} \approx 2.12\lambda$. The phase-encoded input wave $\exp(j\psi)$ propagates a distance $d_{to} \approx 13.33\lambda$ to the plane of the opaque occlusion, where its amplitude is modulated by the occlusion function $o(x, y)$ such that:

$$o(x,y) = \begin{cases} 0, & |x| < \frac{w_o}{2}, |y| < \frac{w_o}{2} \\ 1, & \text{otherwise} \end{cases}$$

The encoded wave, after being obstructed and scattered by the occlusion, travels to the receiver through free space. At the receiver, the diffractive decoder all-optically processes and decodes the incoming wave to produce an all-optical reconstruction $\hat{m}'$ of the original message $m$ at its output FOV. We assume the receiver aperture, which coincides with the first layer of the diffractive decoder, to be located at an axial distance of $d_{ol} \approx 106.67\lambda$ away from the plane of the occlusion. The effective size of the independent diffractive features of each transmissive layer is assumed to be $0.53\lambda \times 0.53\lambda$, and each of the $L$ layers comprises $200 \times 200$ such diffractive features, resulting in a lateral width of $w_l \approx$ $106.67\lambda$ for the diffractive layers. The layer-to-layer separation is assumed to be $d_{ll} = 40\lambda$. The output FOV of the diffractive decoder is assumed to be $40\lambda$ away from the last diffractive layer and extend over an area $w_d \times w_d$, where $w_d \approx 59.73\lambda$.

The diffractive decoding at the receiver involves consecutive modulation of the received wave by the $L$ diffractive layers, each followed by propagation through the free space. The modulation of the incident optical wave on a diffractive layer is assumed to be realized passively by its height variations. The complex transmittance $\tilde{t}(x, y)$ of a passive diffractive layer is related to its height $h(x, y)$ according to:

$$\tilde{t} = \exp\left(j\frac{2\pi}{\lambda}(n + jk - 1)h\right) = \exp\left(-\frac{2\pi k}{\lambda}h\right)\exp\left(j\frac{2\pi}{\lambda}(n - 1)h\right) = a\exp(j\varphi)$$

where $n$ and $k$ are the refractive index and the extinction coefficient, respectively, of the diffractive layer material at $\lambda$; $a = \exp\left(-\frac{2\pi k}{\lambda}h\right)$ and $\varphi = \frac{2\pi}{\lambda}(n - 1)h$ are the amplitude and the phase of the complex field transmittance, respectively. For our numerical simulations, we assume the diffractive layers to be lossless, i.e., $k = 0$, $a = 1$, unless stated otherwise.

The propagation of the optical fields through free space is modeled using the angular spectrum method (*33*, *42*), according to which the transformation of an optical field $u(x, y)$ after propagation by an axial distance $d$ can be computed as follows:

$$u(x, y; z = z_0 + d) = \mathcal{F}^{-1}\{\mathcal{F}\{u(x, y; z = z_0)\} \times H(f_x, f_y; d)\}$$

where $\mathcal{F}$ ($\mathcal{F}^{-1}$) is the two-dimensional Fourier (Inverse Fourier) transform operator and $H(f_x, f_y; d)$ is the free-space transfer function for propagation by an axial distance $d$ defined as follows:

$$H(f_x, f_y; d) = \begin{cases} \exp\left(j\frac{2\pi}{\lambda}d\sqrt{1 - (\lambda f_x)^2 - (\lambda f_y)^2}\right), & f_x^2 + f_y^2 < 1/\lambda^2 \\ 0, & \text{otherwise} \end{cases}$$

In our numerical analyses, the optical fields were sampled at an interval of $\delta \approx 0.53\lambda$ along both $x$ and $y$ directions and the Fourier (Inverse Fourier) transforms were implemented using the Fast Fourier Transform (FFT) algorithm.

For the lens-based imaging simulations reported in this work, the plane wave illumination was assumed to be amplitude modulated by the object placed at the transmitter aperture, and the (thin) lens is assumed to be placed at the same plane as the plane of the first diffractive layer in the encoding-decoding scheme, with the diameter of the lens aperture equal to the width of the diffractive layer, i.e., $w_l \approx 106.67\lambda$.

## Experimental design

In our experiments, the wavelength of operation was $\lambda = 0.75\text{mm}$. We used a single-layer diffractive decoder, i.e., $L = 1$, with $N = 200^2$ independent features and the width of each feature was $\sim 0.53\lambda \approx 0.40\text{mm}$, resulting in an $\sim 80\text{mm} \times 80\text{mm}$ diffractive layer. The width of the transmitter aperture accommodating the encoded phase messages was $w_t \approx 59.73\lambda \approx 44.8\text{mm}$, same as the width of the output FOV $w_d$. The occlusion width was $w_o \approx 32\lambda \approx 24\text{mm}$. The distance from the transmitter aperture to the occlusion plane was $d_{to} \approx 13.33\lambda \approx 10\text{mm}$, while the diffractive layer was $d_{ol} \approx$ $106.67\lambda \approx 80$ mm away from the occlusion plane. The output FOV was $40\lambda \approx 30$mm away from the diffractive layer.

The diffractive layers and the phase-encoded messages (CNN outputs) were fabricated using a 3D printer (Objet30 Pro, Stratasys Ltd). Similar to the implementation of the diffractive layer phase, the phase-encoded messages were implemented by height variations according to $h_o = \psi \frac{\lambda}{2\pi(n-1)}$. The height variations were applied on top of a uniform base thickness of 0.2mm, used for mechanical support. The occlusion was realized by pasting aluminum on a 3D-printed substrate (see Fig. 9). The measured complex refractive index $n + jk$ of the 3D-printing material at $\lambda = 0.75$mm was $1.6518 + j0.0612$.

While training the experimental model, the weight $\eta$ of the diffraction efficiency-related loss term was set to be zero. To make the experimental design robust against misalignments, we incorporated random lateral and axial misalignments of the encoded objects, the occlusion and the diffractive layer into the optical forward model during its training (*45*). The random misalignments were modeled using the uniformly distributed random variables $\Delta_x \sim Uniform(-\lambda, \lambda)$, $\Delta_y \sim Uniform(-\lambda, \lambda)$ and $\Delta_z \sim Uniform(-2\lambda, 2\lambda)$ representing the displacements of the encoded objects, the occlusion and the diffractive layer along $x$, $y$ and $z$ directions, respectively, from their nominal positions.

## Terahertz experimental setup

A WR2.2 modular amplifier/multiplier chain (AMC) in conjunction with a compatible diagonal horn antenna from Virginia Diodes Inc. was used to generate a continuous-wave (CW) radiation at 0.4 THz, by multiplying a 10 dBm RF input signal at $f_{RF1}$ = 11.1111 GHz 36 times. To resolve low-noise output data through lock-in detection, the AMC output was modulated at a rate of $f_{MOD}$ = 1 kHz. The exit aperture of the horn antenna was positioned ~60 cm away from the input (encoded object) plane of the 3D-printed diffractive decoder for the incident THz wavefront to be approximately planar. A single-pixel Mixer/AMC, also from Virginia Diodes Inc., was used to detect the diffracted THz radiation at the output plane. To down-convert the detected signal to 1 GHz, a 10dBm local oscillator signal at $f_{RF1}$ = 11.0833 GHz was fed to the detector. The detector was placed on an X-Y positioning stage consisting of two linear motorized stages from Thorlabs NRT100, and the output FOV was scanned using a 0.5 × 0.1 mm detector with a scanning interval of 2 mm. The down-converted signal was amplified, using cascaded low-noise amplifiers from Mini-Circuits ZRL-1150-LN+, by 40 dB and passed through a 1 GHz (+/-10 MHz) bandpass filter (KL Electronics 3C40-1000/T10-O/O) to filter out the noise from unwanted frequency bands. The filtered signal was attenuated by a tunable attenuator (HP 8495B) for linear calibration and then detected by a low-noise power detector (Mini-Circuits ZX47-60). The output voltage signal was read out using a lock-in amplifier (Stanford Research SR830), where the $f_{MOD}$ = 1kHz modulation signal served as the reference signal. The lock-in amplifier readings were converted to a linear scale according to the calibration results. To enhance the signal-to-noise ratio (SNR), a $2 \times 2$ binning was applied to the THz measurements. We also digitally enhanced the contrast of the measurements by saturating the top 1% and the bottom 1% of the pixel values using the built-in MATLAB function *imadjust* and mapping the resulting image to a dynamic range between 0 and 1.

**Supplementary Materials:** This file contains the training details and Supplementary Figures S1-S4.

## References

1. D. O'Brien, G. Parry, P. Stavrinou, Optical hotspots speed up wireless communication. *Nat. Photonics*. **1**, 245–247 (2007).

2. P. Yang, Y. Xiao, M. Xiao, S. Li, 6G Wireless Communications: Vision and Potential Techniques. *IEEE Netw.* **33**, 70–75 (2019).

3. Z. Xu, B. M. Sadler, Ultraviolet Communications: Potential and State-Of-The-Art. *IEEE Commun. Mag.* **46**, 67–73 (2008).

4. X. He, E. Xie, M. S. Islim, A. A. Purwita, J. J. D. McKendry, E. Gu, H. Haas, M. D. Dawson, 1 Gbps free-space deep-ultraviolet communications based on III-nitride micro-LEDs emitting at 262 nm. *Photonics Res.* **7**, B41–B47 (2019).

5. C. H. Kang, I. Dursun, G. Liu, L. Sinatra, X. Sun, M. Kong, J. Pan, P. Maity, E.-N. Ooi, T. K. Ng, O. F. Mohammed, O. M. Bakr, B. S. Ooi, High-speed colour-converting photodetector with all-inorganic CsPbBr3 perovskite nanocrystals for ultraviolet light communication. *Light Sci. Appl.* **8**, 94 (2019).

6. S. Arnon, D. Kedar, Non-line-of-sight underwater optical wireless communication network. *JOSA A*. **26**, 530–539 (2009).

7. L. Wang, Z. Xu, B. M. Sadler, An approximate closed-form link loss model for non-line-of-sight ultraviolet communication in noncoplanar geometry. *Opt. Lett.* **36**, 1224–1226 (2011).

8. H. Xiao, Y. Zuo, J. Wu, Y. Li, J. Lin, Non-line-of-sight ultraviolet single-scatter propagation model in random turbulent medium. *Opt. Lett.* **38**, 3366–3369 (2013).

9. Z. Cao, X. Zhang, G. Osnabrugge, J. Li, I. M. Vellekoop, A. M. J. Koonen, Reconfigurable beam system for non-line-of-sight free-space optical communication. *Light Sci. Appl.* **8**, 69 (2019).

10. Z. Liu, Y. Huang, H. Liu, X. Chen, Non-line-of-sight optical communication based on orbital angular momentum. *Opt. Lett.* **46**, 5112–5115 (2021).

11. A. Velten, T. Willwacher, O. Gupta, A. Veeraraghavan, M. G. Bawendi, R. Raskar, Recovering three-dimensional shape around a corner using ultrafast time-of-flight imaging. *Nat. Commun.* **3**, 745 (2012).

12. G. Gariepy, F. Tonolini, R. Henderson, J. Leach, D. Faccio, Detection and tracking of moving objects hidden from view. *Nat. Photonics*. **10**, 23–26 (2016).

13. M. O'Toole, D. B. Lindell, G. Wetzstein, Confocal non-line-of-sight imaging based on the light-cone transform. *Nature*. **555**, 338–341 (2018).

14. C. Saunders, J. Murray-Bruce, V. K. Goyal, Computational periscopy with an ordinary digital camera. *Nature*. **565**, 472–475 (2019).

15. T. Maeda, Y. Wang, R. Raskar, A. Kadambi, "Thermal Non-Line-of-Sight Imaging" in *2019 IEEE International Conference on Computational Photography (ICCP)* (2019), pp. 1–11.

16. F. Heide, M. O'Toole, K. Zang, D. B. Lindell, S. Diamond, G. Wetzstein, *ACM Trans. Graph.*, in press, doi:10.1145/3269977.

17. D. B. Lindell, G. Wetzstein, M. O'Toole, *ACM Trans. Graph.*, in press, doi:10.1145/3306346.3322937.

18. J. Boger-Lombard, O. Katz, Passive optical time-of-flight for non line-of-sight localization. *Nat. Commun.* **10**, 3343 (2019).

19. X. Liu, I. Guillén, M. La Manna, J. H. Nam, S. A. Reza, T. Huu Le, A. Jarabo, D. Gutierrez, A. Velten, Non-line-of-sight imaging using phasor-field virtual wave optics. *Nature*. **572**, 620–623 (2019).

20. M. Kaga, T. Kushida, T. Takatani, K. Tanaka, T. Funatomi, Y. Mukaigawa, Thermal non-line-of-sight imaging from specular and diffuse reflections. *IPSJ Trans. Comput. Vis. Appl.* **11**, 8 (2019).

21. C. A. Metzler, F. Heide, P. Rangarajan, M. M. Balaji, A. Viswanath, A. Veeraraghavan, R. G. Baraniuk, Deep-inverse correlography: towards real-time high-resolution non-line-of-sight imaging. *Optica*. **7**, 63–71 (2020).

22. X. Liu, S. Bauer, A. Velten, Phasor field diffraction based reconstruction for fast non-line-of-sight imaging systems. *Nat. Commun.* **11**, 1645 (2020).

23. C. Wu, J. Liu, X. Huang, Z.-P. Li, C. Yu, J.-T. Ye, J. Zhang, Q. Zhang, X. Dou, V. K. Goyal, F. Xu, J.-W. Pan, Non–line-of-sight imaging over 1.43 km. *Proc. Natl. Acad. Sci.* **118**, e2024468118 (2021).

24. B. Wang, M.-Y. Zheng, J.-J. Han, X. Huang, X.-P. Xie, F. Xu, Q. Zhang, J.-W. Pan, Non-Line-of-Sight Imaging with Picosecond Temporal Resolution. *Phys. Rev. Lett.* **127**, 053602 (2021).

25. J. Maycock, C. P. McElhinney, B. M. Hennelly, T. J. Naughton, J. B. McDonald, B. Javidi, Reconstruction of partially occluded objects encoded in three-dimensional scenes by using digital holograms. *Appl. Opt.* **45**, 2975–2985 (2006).

26. Z. Yaqoob, D. Psaltis, M. S. Feld, C. Yang, Optical phase conjugation for turbidity suppression in biological samples. *Nat. Photonics*. **2**, 110–115 (2008).

27. Y. Rivenson, A. Rot, S. Balber, A. Stern, J. Rosen, Recovery of partially occluded objects by applying compressive Fresnel holography. *Opt. Lett.* **37**, 1757–1759 (2012).

28. Y. Xiao, L. Zhou, W. Chen, Wavefront control through multi-layer scattering media using single-pixel detector for high-PSNR optical transmission. *Opt. Lasers Eng.* **139**, 106453 (2021).

29. Y. Xiao, L. Zhou, Z. Pan, Y. Cao, W. Chen, Physically-secured high-fidelity free-space optical data transmission through scattering media using dynamic scaling factors. *Opt. Express*. **30**, 8186–8198 (2022).

30. Z. Pan, Y. Xiao, Y. Cao, L. Zhou, W. Chen, Accurate optical information transmission through thick tissues using zero-frequency modulation and single-pixel detection. *Opt. Lasers Eng.* **158**, 107133 (2022).

31. S. Popoff, G. Lerosey, M. Fink, A. C. Boccara, S. Gigan, Image transmission through an opaque material. *Nat. Commun.* **1**, 81 (2010).

32. L. Zhu, F. Soldevila, C. Moretti, A. d'Arco, A. Boniface, X. Shao, H. B. de Aguiar, S. Gigan, Large field-of-view non-invasive imaging through scattering layers using fluctuating random illumination. *Nat. Commun.* **13**, 1447 (2022).

33. X. Lin, Y. Rivenson, N. T. Yardimci, M. Veli, Y. Luo, M. Jarrahi, A. Ozcan, All-optical machine learning using diffractive deep neural networks. *Science*. **361**, 1004–1008 (2018).

34. MNIST handwritten digit database, Yann LeCun, Corinna Cortes and Chris Burges, (available at http://yann.lecun.com/exdb/mnist/).

35. Z. Wang, A. C. Bovik, H. R. Sheikh, E. P. Simoncelli, Image quality assessment: from error visibility to structural similarity. *IEEE Trans. Image Process.* **13**, 600–612 (2004).

36. H. Xiao, K. Rasul, R. Vollgraf, Fashion-MNIST: a Novel Image Dataset for Benchmarking Machine Learning Algorithms. *ArXiv170807747 Cs Stat* (2017) (available at http://arxiv.org/abs/1708.07747).

37. CIFAR-10 and CIFAR-100 datasets, (available at https://www.cs.toronto.edu/~kriz/cifar.html).

38. O. Kulce, D. Mengu, Y. Rivenson, A. Ozcan, All-optical information-processing capacity of diffractive surfaces. *Light Sci. Appl.* **10**, 25 (2021).

39. O. Kulce, D. Mengu, Y. Rivenson, A. Ozcan, All-optical synthesis of an arbitrary linear transformation using diffractive surfaces. *Light Sci. Appl.* **10**, 196 (2021).

40. J. Li, T. Gan, B. Bai, Y. Luo, M. Jarrahi, A. Ozcan, Massively parallel universal linear transformations using a wavelength-multiplexed diffractive optical network. *Adv. Photonics*. **5**, 016003 (2023).

41. M. S. S. Rahman, X. Yang, J. Li, B. Bai, A. Ozcan, Universal Linear Intensity Transformations Using Spatially-Incoherent Diffractive Processors (2023), , doi:10.48550/arXiv.2303.13037.

42. J. W. Goodman, *Introduction to Fourier Optics* (Roberts and Company Publishers, 2005).

43. M. Abadi, P. Barham, J. Chen, Z. Chen, A. Davis, J. Dean, M. Devin, S. Ghemawat, G. Irving, M. Isard, M. Kudlur, J. Levenberg, R. Monga, S. Moore, D. G. Murray, B. Steiner, P. Tucker, V. Vasudevan, P. Warden, M. Wicke, Y. Yu, X. Zheng, "TensorFlow: a system for large-scale machine learning" in *Proceedings of the 12th USENIX conference on Operating Systems Design and Implementation* (USENIX Association, USA, 2016), *OSDI'16*, pp. 265–283.

44. D. P. Kingma, J. Ba, Adam: A Method for Stochastic Optimization (2014) (available at https://arxiv.org/abs/1412.6980v9).

45. D. Mengu, Y. Zhao, N. T. Yardimci, Y. Rivenson, M. Jarrahi, A. Ozcan, Misalignment resilient diffractive optical networks. *Nanophotonics*. **1** (2020), doi:10.1515/nanoph-2020-0291.

## Figures

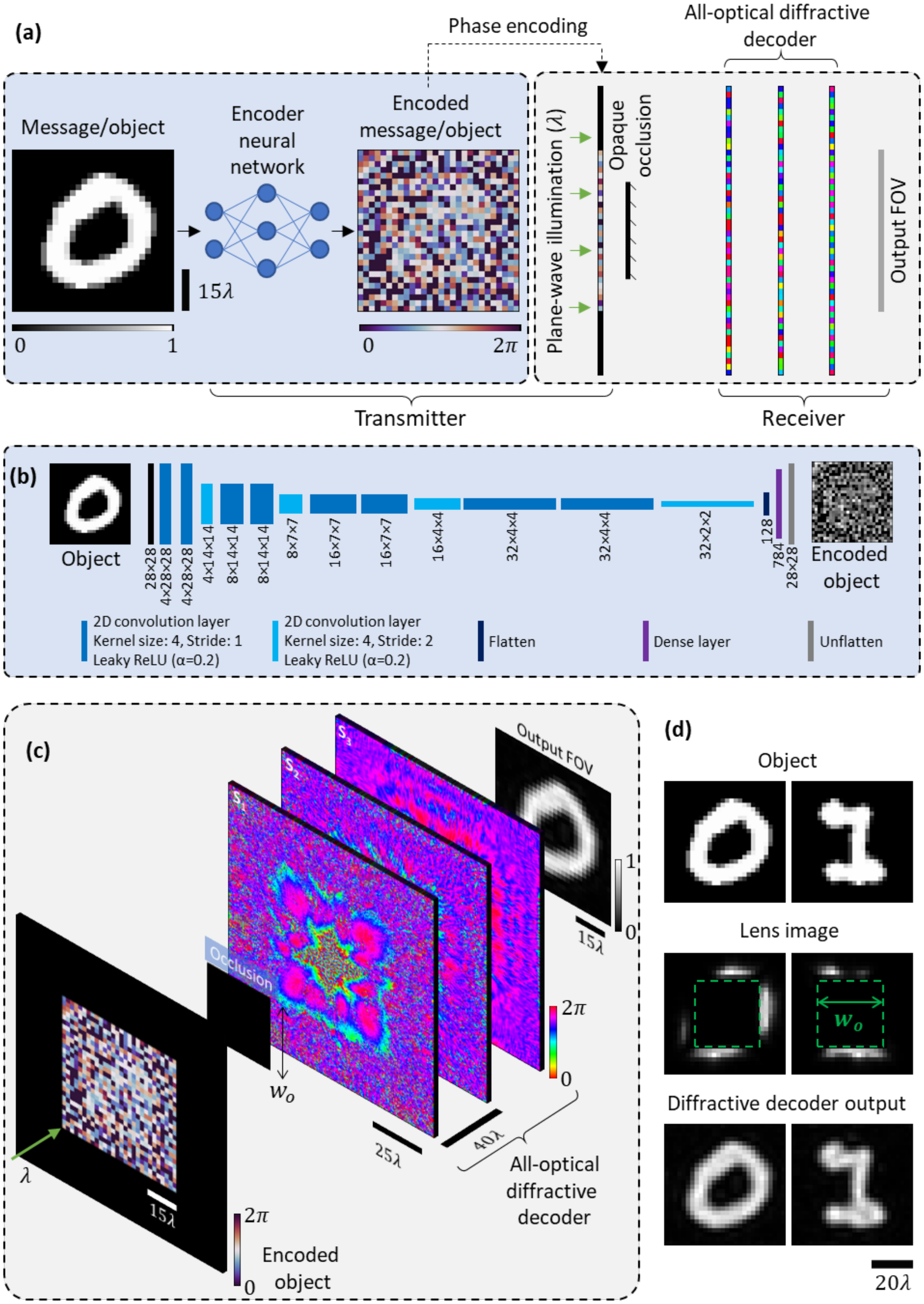

**Fig. 1** Schematic of the optical communication framework around fully opaque occlusions using electronic encoding and diffractive all-optical decoding. (a) An electronic neural network encoder and an all-optical diffractive decoder are trained jointly for communicating around an opaque occlusion. For a message/object to be transmitted, the electronic encoder outputs a coded 2D phase pattern, which is imparted onto a plane wave at the transmitter aperture. The phase-encoded wave, after being obstructed and scattered by the fully opaque occlusion, travels to the receiver, where the diffractive decoder all-optically processes the encoded information to reproduce the message on its output FOV. (b) The architecture used for the convolutional neural network (CNN) encoder throughout this work. (c) Visualization of different processes, such as the obstruction of the transmitted phase-encoded wave by the occlusion of width $w_o$ and the subsequent all-optical decoding performed by the diffractive decoder. The diffractive decoder comprises $L$ surfaces ($S_1, \cdots, S_L$) with phase-only diffractive features. In this figure, $L = 3$ is illustrated as an example. (d) Comparison of the encoding-decoding scheme against conventional lens-based imaging.

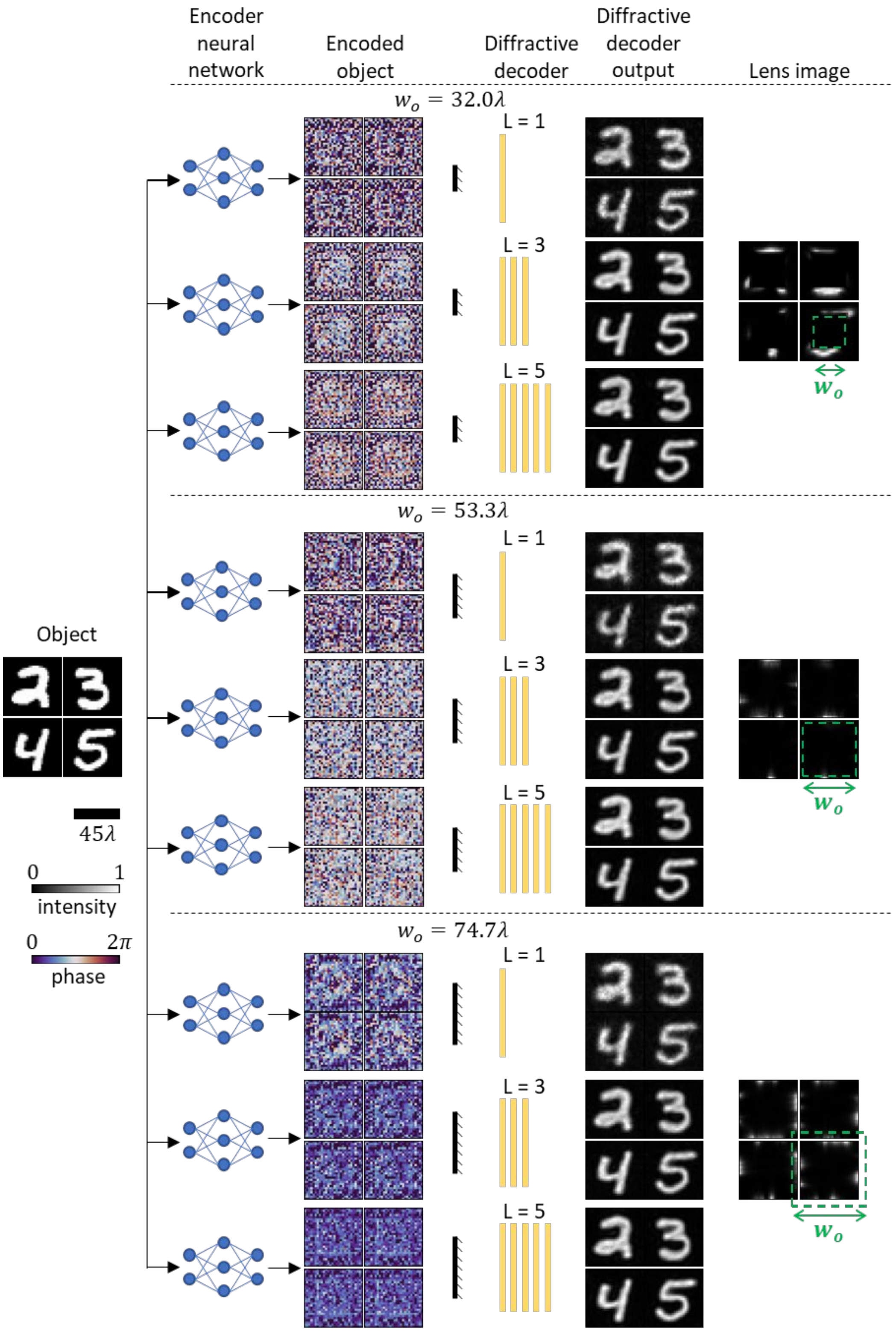

**Fig. 2** Generalization of trained encoder-decoder pairs to previously unseen handwritten digit objects. For different values of the occlusion width $w_o$, the performances of trained encoder-decoder pairs with different numbers of decoder layers ($L$) are depicted for comparison.

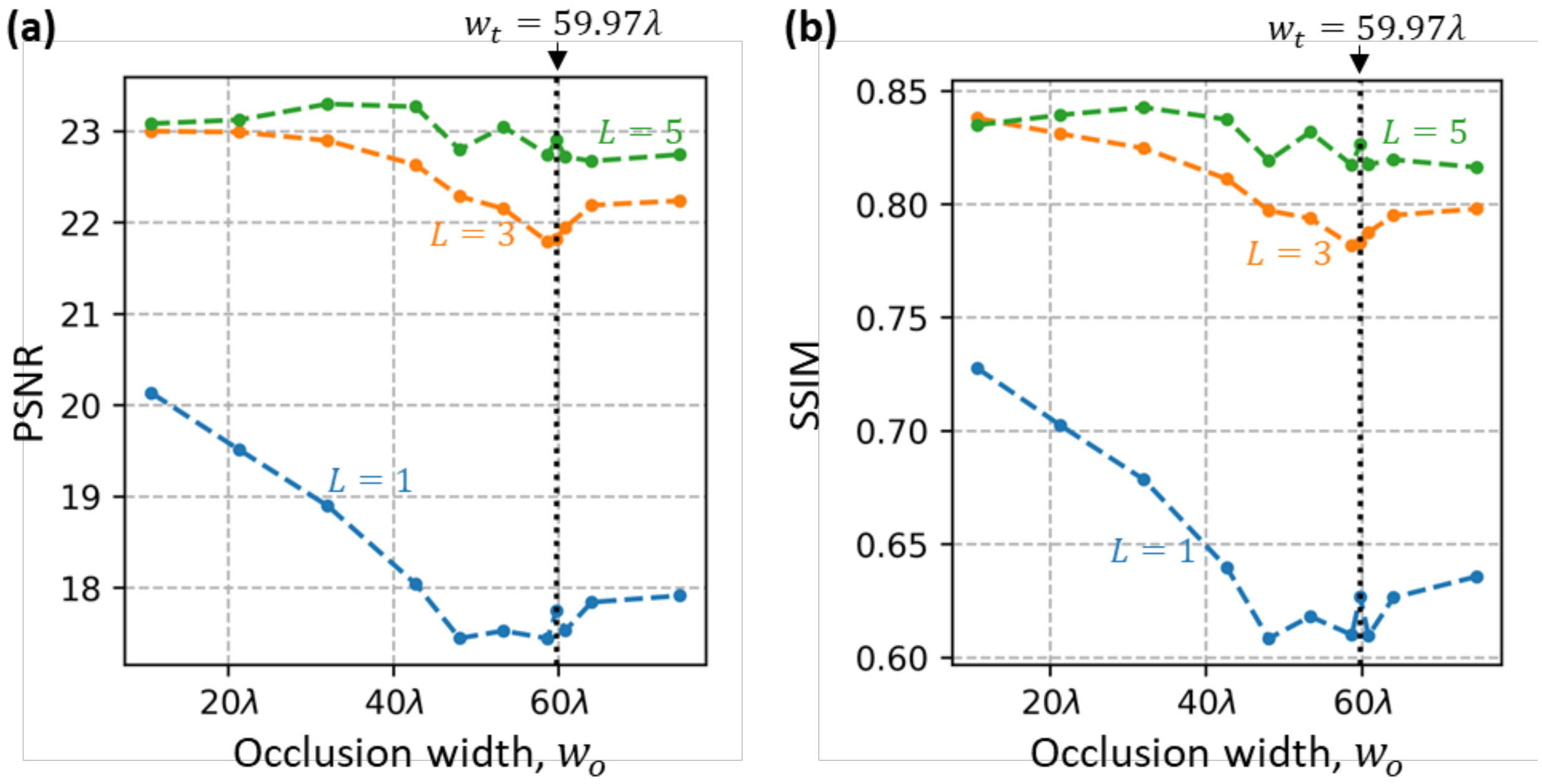

**Fig. 3** Quantification of the performance of encoder-decoder pairs with different numbers of decoder layers ($L$) trained for increasing occlusion widths ($w_o$) in terms of (a) PSNR and (b) SSIM between the diffractive decoder outputs and the ground-truth messages. The PSNR and SSIM values are calculated by averaging over 10,000 MNIST test images. $w_t$ refers to the width of the transmitter aperture.

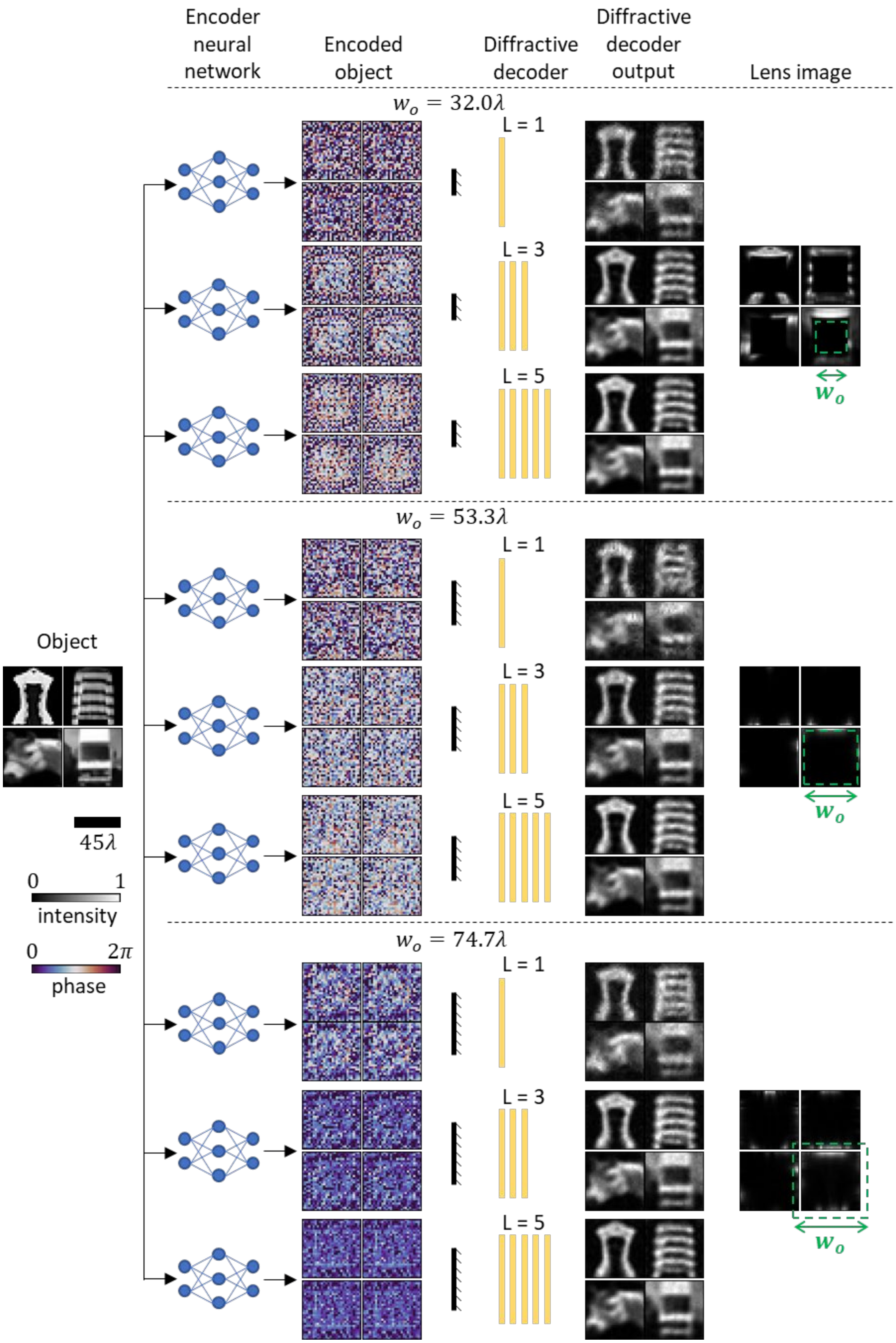

**Fig. 4** Same as Fig. 2, except that these results reflect external generalizations on object types different from those used during the training.

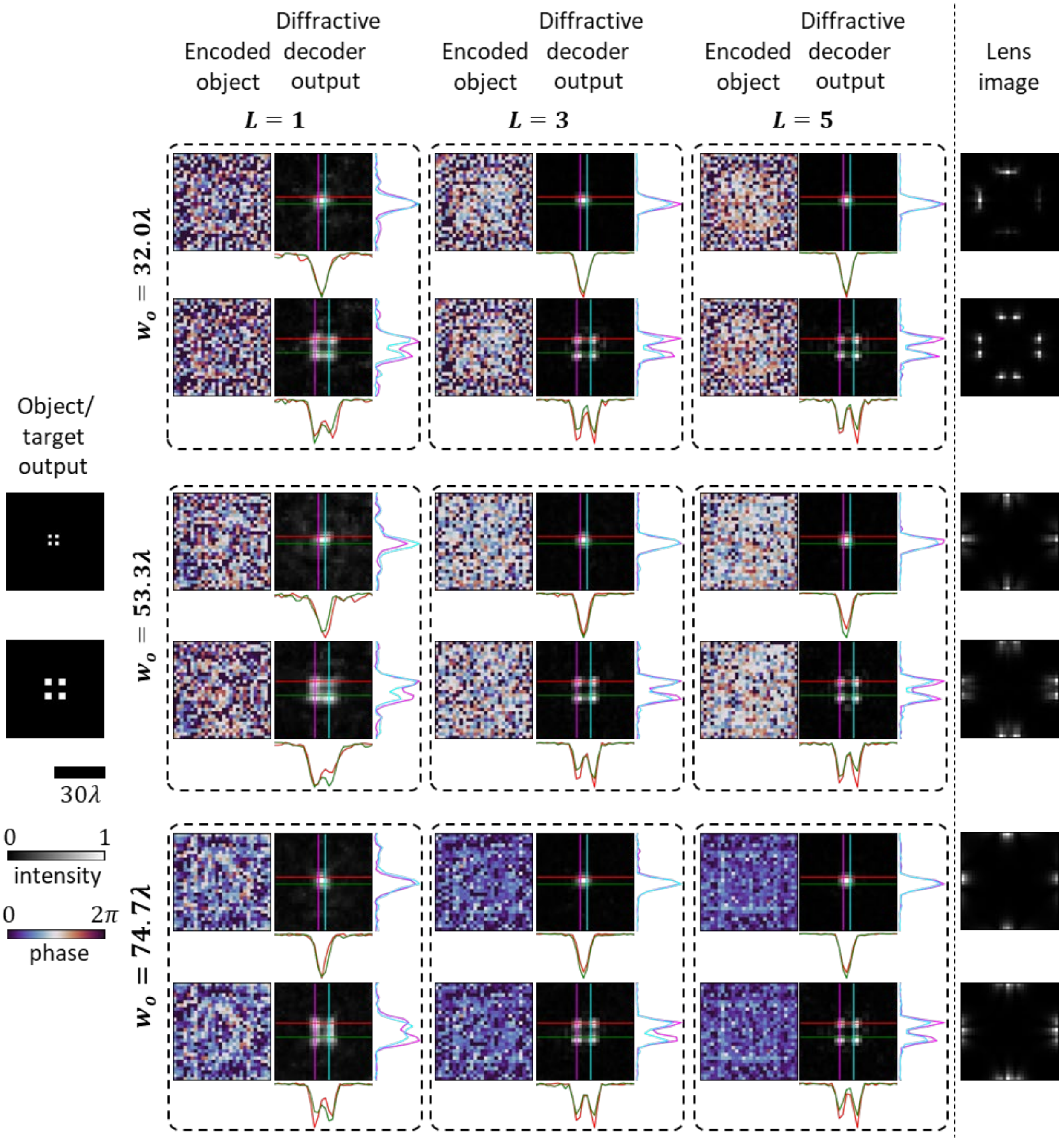

**Fig. 5** Output resolution of diffractive decoders corresponding to $L = 1$, $L = 3$, and $L = 5$ designs trained for different occlusion widths ($w_o$). As for the objects, the vertical/horizontal separation between the inner edges of the dots is $2.12\lambda$ for the test pattern on the top and $4.24\lambda$ for the one below. The diffractive decoder outputs are accompanied by cross-sections taken along the color-coded vertical/horizontal lines.

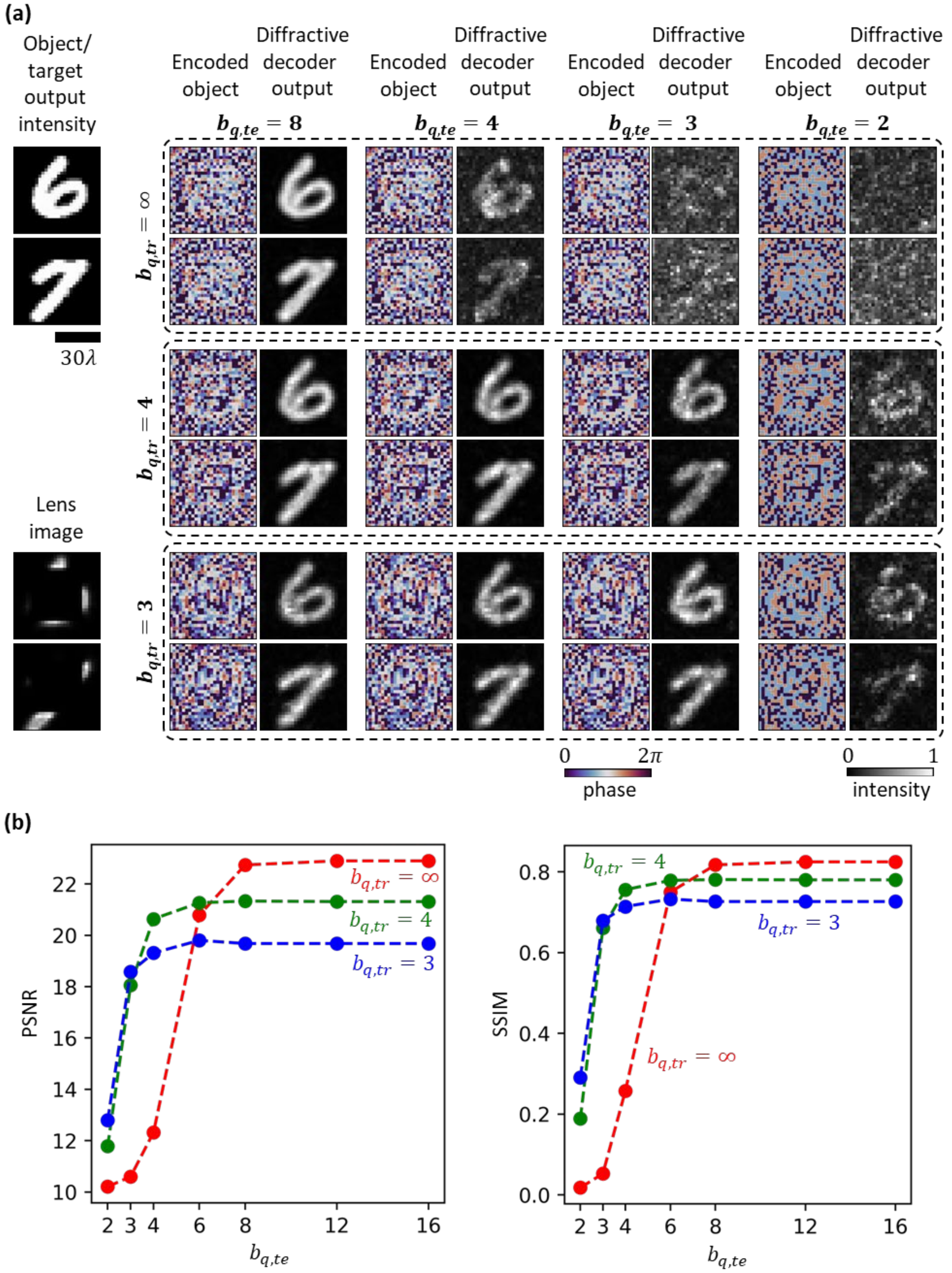

**Fig. 6** Effect of the phase bit depth of the encoded object and the diffractive layer features on the performance of trained encoder-decoder pairs. (a) Qualitative performance of the designs, which are trained assuming a certain phase quantization bit depth $b_{q,tr}$, reported as a function of the bit depth used during testing $b_{q,te}$. (b) For different $b_{q,tr}$, PSNR and SSIM values are plotted as a function of $b_{q,te}$. The PSNR and SSIM values are evaluated by averaging the results of 10,000 test images from the MNIST dataset.

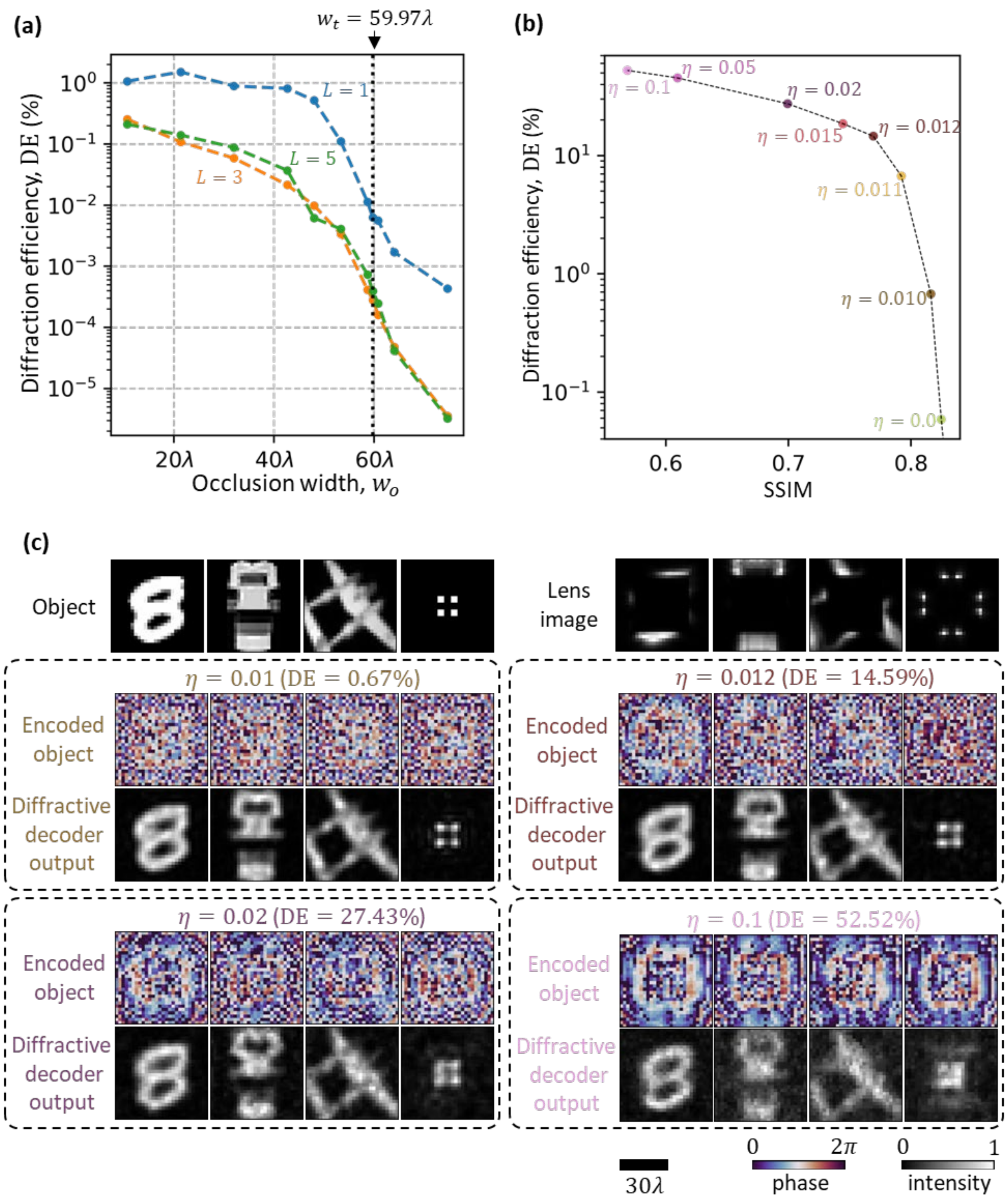

**Fig. 7** Output power efficiency of the electronic encoding-diffractive decoding scheme for optical communication around fully opaque occlusions. (a) Diffraction efficiency (DE) of the same designs shown in Fig. 3. (b) The trade-off between DE and SSIM achieved by varying the training hyperparameter $\eta$, i.e., the weight of an additive loss term used for penalizing low-efficiency designs. For these designs, $w_o = 32\lambda$ and $L = 3$ were used. The DE and SSIM values are calculated by averaging over 10,000 MNIST test images. (c) The performance of some of the designs shown in (b), trained with different $\eta$ values.

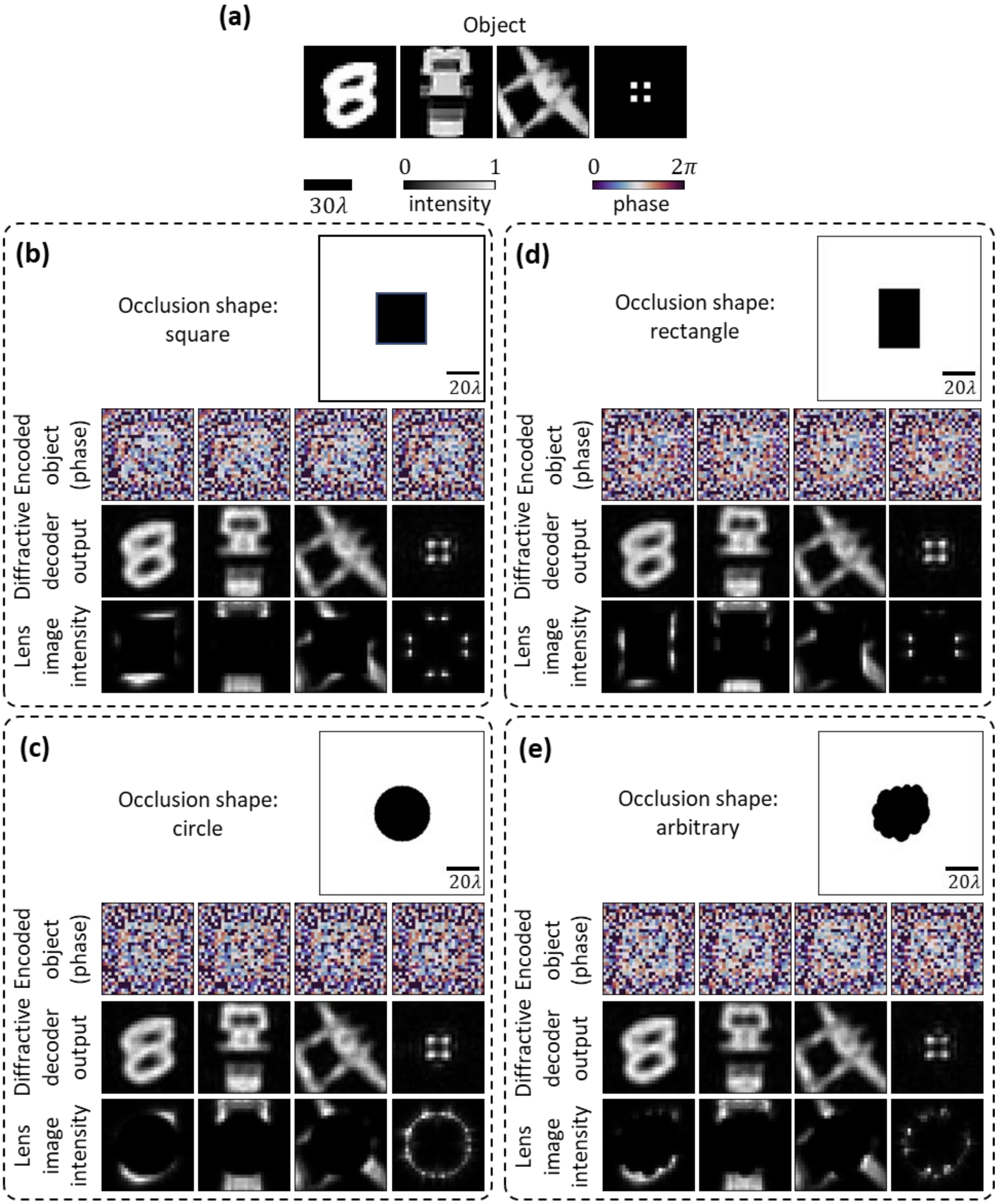

**Fig. 8** Performance of encoder-decoder pairs trained for different opaque occlusion shapes. The performances of four designs trained for different occlusion shapes, i.e., a square, a circle, a rectangle, and an arbitrary shape, are shown. The areas of these fully opaque occlusions are approximately equal.

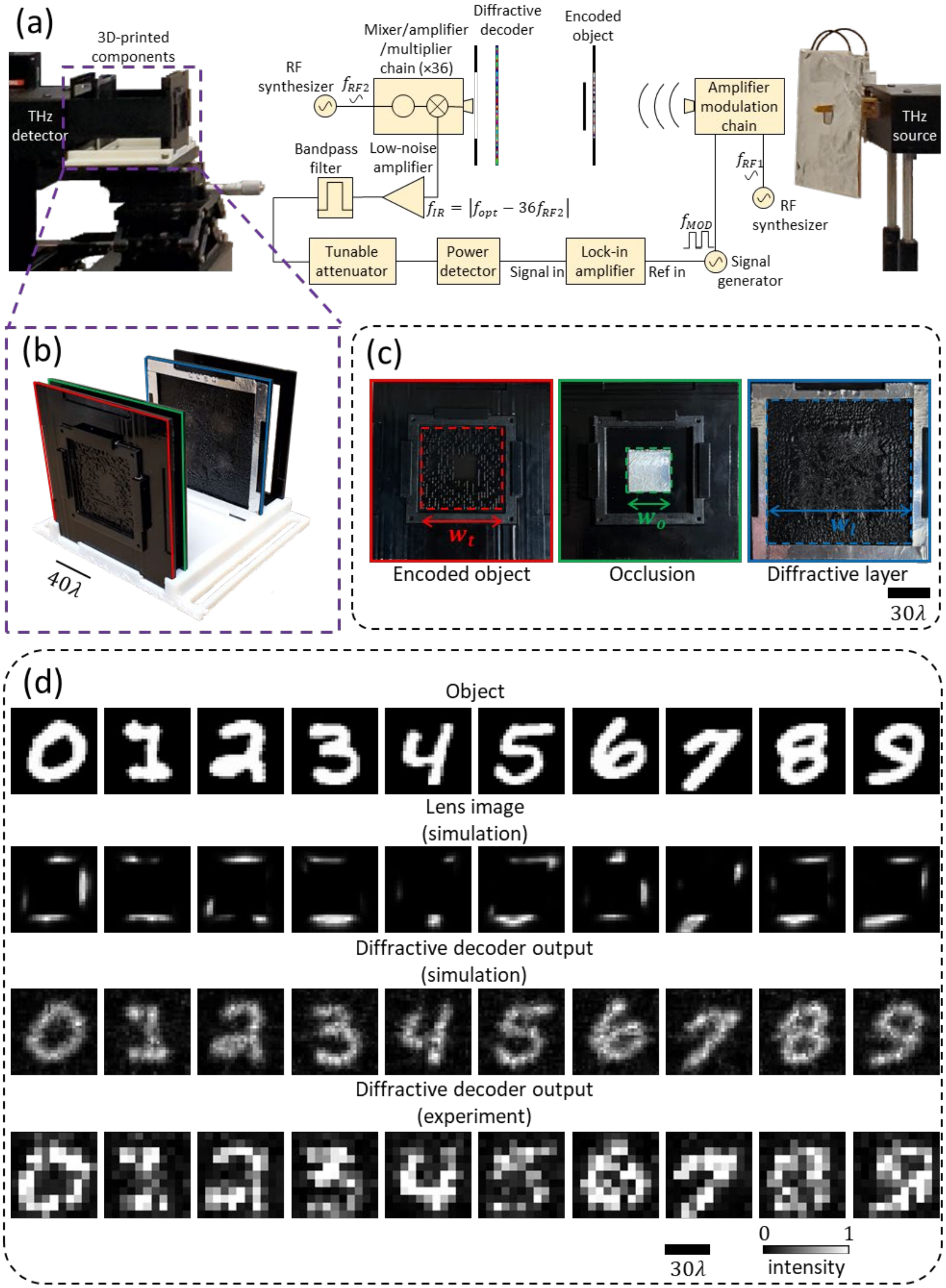

**Fig. 9** Experimental results with an $L = 1$ design for an occlusion width of $w_o = 32\lambda$ operating at a wavelength of $\lambda = 0.75\text{mm}$. (a) The terahertz setup comprising the source and the detector, together with the 3D-printed components used as the encoded phase objects, the occlusion, and the diffractive

layer. (b) Assembly of the encoded phase objects, the occlusion, the diffractive layer, and the output aperture using a 3D-printed holder. (c) The encoded phase object (one example), the occlusion, and the diffractive layer are shown separately, housed inside the supporting frames. (d) Experimental diffractive decoder outputs (bottom row) for ten handwritten digit objects (top row), together with the corresponding simulated lens images (second row) and the diffractive decoder outputs (third row).